\documentclass[aps, 
 prl, 
 reprint, 
 superscriptaddress, 
 floatfix, 
 tightenlines, 
 twocolumn
 ]{revtex4-2}

\usepackage[english]{babel}
\usepackage{graphicx} 
\usepackage{listings} 
\usepackage{amsmath,amssymb,amsthm} 
\usepackage[utf8]{inputenc}
\usepackage[toc,page]{appendix}
\usepackage{color}

\usepackage[export]{adjustbox}

\usepackage{hyperref}
\hypersetup{
  colorlinks = true, 
  urlcolor = blue, 
  linkcolor = blue, 
  citecolor = blue 
}

\newcommand{\ket}[1]{\left | #1 \right \rangle} 
\newcommand{\tx}[1]{\textup{#1}} 
\newcommand{\YSO}{Y$_2$SiO$_5$}
\newcommand{\PrYSO}{Pr$^\textup{3+}$:Y$_2$SiO$_5$}

\begin{document}

\title{Transmission of light-matter entanglement over a metropolitan network}
\author{Jelena V. Rakonjac}
\altaffiliation{These authors contributed equally. Corresponding author: samuele.grandi@icfo.eu}
\author{Samuele Grandi}
\altaffiliation{These authors contributed equally. Corresponding author: samuele.grandi@icfo.eu}
\author{S\"oren Wengerowsky}
\altaffiliation{These authors contributed equally. Corresponding author: samuele.grandi@icfo.eu}
\author{Dario Lago-Rivera}
\author{F\'elicien Appas}
\affiliation{ICFO-Institut de Ciencies Fotoniques, The Barcelona Institute of Science and Technology, 08860 Castelldefels (Barcelona), Spain.}
\author{Hugues de Riedmatten}
\affiliation{ICFO-Institut de Ciencies Fotoniques, The Barcelona Institute of Science and Technology, 08860 Castelldefels (Barcelona), Spain.}
\affiliation{ICREA-Instituci\'o Catalana de Recerca i Estudis Avan\c cats, 08015 Barcelona, Spain.}

\begin{abstract}
We report on the transmission of telecom photons entangled with a multimode solid-state quantum memory over a deployed optical fiber in a metropolitan area. Photon pairs were generated through spontaneous parametric down-conversion, with one photon stored in a rare earth-based quantum memory, and the other, at telecommunication wavelengths, traveling through increasing distances of optical fibre, first in the laboratory and then outside in a deployed fibre loop. We measured highly-non-classical correlations between the stored and the telecom photons for storage times up to 25~$\mu$s and for a fibre separation up to 50~km. We also report light-matter entanglement with a two-qubit fidelity up to 88$\%$, which remains constant within error bars for all fibre lengths, showing that the telecom qubit does not suffer decoherence during the transmission. Finally, we moved the detection stage of the telecom photons to a different location placed 17~km away, and confirmed the non-classical correlations between the two photons. Our system was adapted to provide the transmission of precise detection times and synchronization signals over long quantum communication channels, providing the first steps for a future quantum network involving quantum memories and non-classical states.

\end{abstract}

\maketitle


\section{Introduction}
In standard telecommunications, erbium-doped amplifiers are placed along the network to regenerate the depleted signals, compensating for the attenuation of the optical fibres. The quantum repeater has been introduced in quantum communication \cite{Briegel1998, Duan2001, Sangouard2011} to solve the same problem of attenuation, since the amplification of quantum states is not possible without a critical decrease in the qubit fidelity due to the no-cloning theorem, and therefore enable long-distance quantum networks, offering new opportunities in quantum communication, computing and sensing \cite{Wehner2018}. One of the main challenges of building such a quantum network is to efficiently deliver entanglement between the remote nodes \cite{Kimble2008}. Entanglement is the main resource in quantum communication, as it can be used e.g. to perform quantum key distribution or to transmit qubits between nodes using quantum teleportation \cite{Bennett1993}. Similarly, it is at the base of the operation of quantum repeaters, which allow long-distance entanglement distribution through successive entanglement swapping operations combined with quantum memories.

Any system addressing long-distance quantum communication has to overcome several hurdles to achieve even the first stepping stone, namely quantum correlations between a matter qubit and a photon. While several different protocols have been proposed \cite{Duan2001,Sangouard2011,Sinclair2014,Borregaard2020}, all involve an information carrier propagating through several kilometres of communication channels. The commercial telecommunication network would then provide a natural platform over which to implement early tests of quantum communication. However, only a few quantum systems can operate at telecom wavelength, and accessing it with quantum frequency conversion increases system losses and may reduce signal-to-noise. A promising option involves combining a quantum memory with a non-degenerate source of photon pairs, where one photon is compatible with the memory, while the other is at telecom wavelength \cite{Simon2007}. Another desirable feature for a functional quantum repeater is the use of multiplexed communication \cite{Simon2007}. It requires the encoding, storage and transmission of quantum information over several degrees of freedom, such as time, frequency and space, each holding as many modes as possible. While this technique is routinely used in commercial networks, adapting it to the quantum regime is a task only recently faced in entanglement distribution, especially with quantum memories \cite{deRiedmatten2008,Afzelius2009,Sinclair2014,Pu2017,Parniak2017,Tian2017,Seri2019,Li2022}.

Several architectures are currently being investigated for the realisation of quantum repeaters, based on the interaction of photons with individual quantum systems, such as ions and atoms \cite{Moehring2007, Ritter2012, Hofmann2012, Daiss2021, Krutyanskiy2022}, or with atomic ensembles \cite{Chou2005, Chou2007, Farrera2016a, Cao2020, Li2020b, Li2020a, Li2021}. Solid state equivalents to the former, such as quantum dots \cite{Delteil2016, Stockill2017} or colour centres in diamond \cite{Bernien2013, Hensen2015, Sipahigil2016, Tchebotareva2019, Pompili2021, Stolk2022}, are also being investigated for their scalability potential. Distribution of entanglement between systems linked by tens of kilometres of optical fibre have been demonstrated in some of the previous systems. Light-matter entanglement between a telecom photon and a single ion was demonstrated after propagation in 50~km of optical fibre \cite{Krutyanskiy2019}. Entanglement between two adjacent atomic clouds was recently demonstrated \cite{Yu2020} where the heralding signal travelled through tens of kilometres of deployed fibre. This was followed by the distribution of light-matter entanglement between two quantum memories physically separated by 12.5 km \cite{Luo2022}. More recently, the heralded entanglement between two single atoms separated by 33~km of optical fiber \cite{vanLeent2022} was also demonstrated.

Rare earth-doped crystals can be considered a solid-state version of an atomic ensemble, with billions of ions trapped inside a crystalline matrix \cite{Macfarlane2002}. They have long been used as a powerful platform for light-matter interaction. Cooling them to a few kelvin can result in long coherence times of the level of hours \cite{Zhong2015}, a record amongst all of the aforementioned systems. They also possess a great potential for massive multiplexing, as the time, frequency and space degrees of freedom can be harnessed for quantum communication \cite{Ortu2022b}. These properties make them promising systems for quantum repeater applications. In that direction, several important steps have been performed, including the storage of single photons \cite{Saglamyurek2012, Clausen2012, Rielander2014, Saglamyurek2015}, of entanglement \cite{Clausen2011, Saglamyurek2011, Ferguson2016, Rakonjac2021} and the realisation of entangled states between remote memories \cite{Usmani2012,Puigibert2020,Lago-Rivera2021,Liu2021}. More recently, the distribution of non-classical correlations \cite{Businger2022} and entanglement \cite{Lago-Rivera2023} across kilometres of optical fibre was also demonstrated. An important validation step towards the realization of quantum networks with this technology is to move from laboratory to field demonstrations using the installed fiber network. This requires the development of robust quantum hardware operating outside of laboratory environment, and sharing classical information, like synchronisation signals and optical references, between the nodes.

\begin{figure*}[t!]
    \centering
    \includegraphics[width=0.9\textwidth]{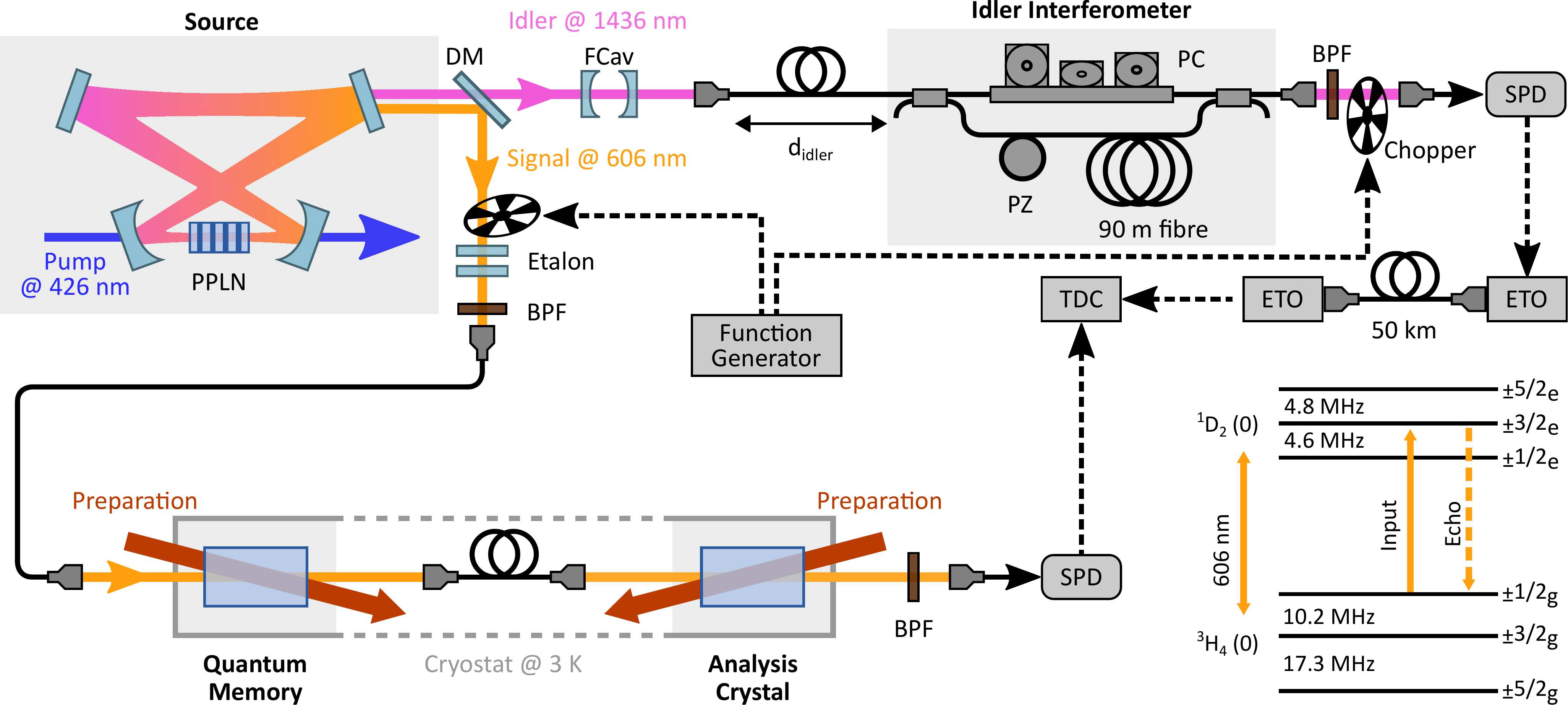}
    \caption{Setup of the experiment. Entangled photon pairs are generated in a cavity-enhanced spontaneous parametric down-conversion, where a periodically-poled lithium niobate (ppLN) crystal is pumped by a laser at 426~nm. The two photons of the pair are separated at a dichroic mirror (DM) separates the signal and idler photons. The idler photons pass through a filter cavity (FCav) for single frequency mode operation and are then fibre-coupled. Signal photons are instead filtered with a band-pass filter (BPF) and an etalon filter before being fibre-coupled and then stored in the quantum memory. The detection signal from the idler detectors are encoded to light and back to an electrical signal using electrical-to-optical (ETO) transducers. The fibre-based idler interforemeter was used only for entanglement verification. TDC: time-to-digital converter. PZ: fibre stretcher. PC: polarisation controller. Bottom right: energy level scheme of praseodymium with the relevant transition identified.}
    \label{fig:setup}
\end{figure*}

In this manuscript we present a set of tests over a quantum network testbed, with the generation of entanglement between a multimode quantum memory based on a rare earth-doped crystal and a telecom photon and its transmission through up to 50~km of deployed optical fibres in the metropolitan area of Barcelona. We show that non-classical correlations and light-matter entanglement are maintained after the transmission, with any degradation coming only from the reduced signal-to-noise ratio. Finally, we fully decouple the photon generation from the detection by realising a transportable detection setup which, placed at another location, we use to demonstrate non-classical correlations between two locations separated by 20 km.

\section{Description of the Setup}

Our design for a quantum network node is based on the combination of a rare earth-doped quantum memory and a photon-pair source \cite{Simon2007}. We employ a crystal of \YSO\ doped with praseodymium (Pr) ions, where we implement the atomic frequency comb (AFC) \cite{Afzelius2009} scheme on the $^3$H$_4$(0) $\leftrightarrow$ $^1$D$_2$(0) transition. This storage protocol is based on the preparation of a periodic absorption profile in the inhomogeneous broadening of an atomic transition, and it has been implemented in bulk crystals \cite{deRiedmatten2008, Bonarota2010, Afzelius2010a, Lauritzen2011, Timoney2012, Businger2022}, erbium-doped fibres \cite{Saglamyurek2014}, waveguides \cite{Saglamyurek2011, Zhong2017, Seri2018, Askarani2019, Liu2020b, Craiciu2021, Rakonjac2022, Su2022}, thin films \cite{Dutta2021} and also warm atomic vapours \cite{Main2021}. Light absorbed by the comb is mapped to a delocalised excitation of the ions in the crystal, each component with a different phase term. However, due to the frequency periodicity $\Delta$ of the comb, these components rephase at time $\tau = 1/\Delta$, causing a collective re-emission in the forward direction. The AFC protocol allows multiplexing in the temporal domain, as many consecutive photons can be stored in the memory, and will be re-emitted in the same order \cite{Afzelius2009}.

The bandwidth of the Pr quantum memory requires a purposely-built source of entangled light. The source we employ is a cavity-enhanced spontaneous parametric down-conversion (SPDC), generating pairs of photon entangled in energy-time \cite{Fekete2013,Seri2018}. The signal and idler photons are generated simultaneously, but where the exact time of generation is distributed within the coherence time of the pump laser of the SPDC. The cavity shapes the spectrum of the photon pair to 1.8~MHz, compatible with the 4~MHz bandwidth of the quantum memory. Moreover, the source is widely non-degenerate, with one photon of the pair, the signal, at 606~nm and resonant with the $^3$H$_4$(0) $\leftrightarrow$ $^1$D$_2$(0) transition of \PrYSO, and the other, the idler, at 1436~nm in the telecom E band. As the spectrum of the photon pair has several frequency modes \cite{Rielander2018}, a Fabry-Perot filter cavity is used to select only one idler mode, with the quantum memory performing a similar filtering on the signal photons. After the source, the signal and idler photons are coupled in optical fibers. The signal photon is stored in the quantum memory for a duration $\tau$, mapping the entanglement of the photon pair onto a light-matter entangled state. After retrieval from the quantum memory, the light is fibre-coupled to clean the spatial mode and passes through another Pr-doped crystal, the analysis crystal, where a transparency window was prepared to act as an ultra---narrow-band filter 4 MHz-wide \cite{Seri2017, Rakonjac2021}.

With this system it is possible to generate quantum correlations and entanglement between a telecom photon and a delocalised excitation stored in a multimode solid-state quantum memory \cite{Rakonjac2021}. We have now tested this system in a practical scenario, where the telecom photons travel through increasing distances $d_\tx{idler}$ of optical fibre (in spools and deployed), with the goal of detecting any deterioration in the non-classical correlations after the transmission. A detailed setup is presented in Fig.~\ref{fig:setup}. Initially the idler photons remain in the laboratory, traveling through only a few metres of fibre at first and then through a 11.4~km fibre spool instead. Next, we sent the idler photons through a dark fibre of the local network, about 25~km long, that connected the ICFO building (in the municipality of Castelldefels) with the Centre for Telecommunication and Technologies of Information (CTTI) in Hospitalet de Llobregat. We then connected it to a second fibre parallel to the first one, such that the photons could be detected back at ICFO after a round-trip $d_\tx{idler} \approx 50$~km and a total loss of $\sim$15~dB at 1436~nm. After transmission through the fibre channel, the idler photons were then coupled back to free space where a band-pass filter removed any noise coming from the metropolitan fibre links. The photons were then fibre-coupled again and sent to superconducting detectors, featuring 85\%\ efficiency and 10~counts/s of background. We used a mechanical chopper to protect the superconducting detectors from the classical telecom light, with the same frequency of the idler photons, which was generated during the locking of the SPDC cavity \cite{Lago-Rivera2021,Rakonjac2021}. A similar chopper was placed on the signal path to block the light used to lock the cavity from reaching the quantum memory setup. The choppers were all synchronised to a 30~Hz signal from a function generator.

Aside from the additional transmission losses of the idler channel, a physical separation between the signal and idler requires of a way of transmitting the detection signal from one position to the other with low jitter. While GPS and ethernet synchronisation is possible \cite{Stolk2022}, an alternative solution is the use of an electrical-to-optical transducer
, turning the electrical signal from the detector to an optical one and then back to electrical at the final destination. We employed this method whenever we included additional distance for the idler photons by placing a 50~km fibre spool between the sender and the receiving transducer. In this manner, the TTL signal of the detector was delayed by $\sim$250~$\mu$s before being recorded by a Time-To-Digital Converter (TDC), therefore simulating distant detection.

\section{Results}
We first tested for any degradation in the quantum correlations between signal and idler by measuring the second-order cross-correlation function $g^{(2)}_\textup{i,AFC} \left ( t \right ) = \frac{p_\tx{i,AFC}}{p_\tx{i}\, p_\tx{s}}$, where $p_\tx{i,AFC}$ is the probability of detecting an idler-signal coincidence at the time of re-emission of the AFC echo and $p_\tx{i}$ and $p_\tx{s}$ are the probabilities of detecting a photon in idler or in the signal mode, respectively, all within the same time window $\Delta t$, which corresponds to the size of the photonic mode. Photon pairs generated by the SPDC source were split at a dichroic mirror, with signal photons stored for a predetermined time in the quantum memory while idler photons were sent through increasing lengths $d_\tx{idler}$. For each one we tested the system by preparing an AFC of $\tau = 10\, \mu$s and $\tau = 25\, \mu$s, with a storage efficiency of 22(1)\%\ and 7(1)\%\ and storing 25 and 62 temporal modes, respectively, considering a mode duration of 400~ns which contains $\sim 92\%$ of the photons.

It should be noted that whenever the idler photon was delayed an additional synchronisation was necessary. When there is no significant distance between the signal and idler photons detection, the pump laser of the SPDC is turned off after the detection of an idler photon to remove the accidental counts from the signal path at the time of AFC emission and to ensure a high signal-to-noise ratio \cite{Seri2017, Rakonjac2021}. At the same time, the TDC stops recording idler detection events, since none would have been emitted and any detection would originate from the background.
However, a delay in the idler path longer than the AFC storage time forbids the use of this technique, as is the case for $d_\tx{idler}$ of 10 and 50~km, where the delays corresponding to the travel times were 50~$\mu$s and 250~$\mu$s, respectively. Therefore, we periodically switched the pump laser on and off for an equal time and with a period corresponding to twice the storage time of the quantum memory. This allowed the storage of the maximum number of temporal modes and their re-emission in a noise-free time window. This generates a periodic noise pattern in the cross-correlation histogram, with a triangular shape coming from the convolution of the two square periodic distributions of the generation of signal and idler photons. We then set the TDC to periodically record idlers for a time equal to the memory storage time, synchronised with the pump laser gating but with a constant offset. With a proper calibration of this offset we could match the recording window of the TDC with the idler generation one. This ensured a maximum signal-to-noise ratio, and as a result placing the AFC echo at the minimum of the triangular noise pattern in the histogram.
\begin{figure}[t]
    \centering
    \includegraphics[width=\columnwidth]{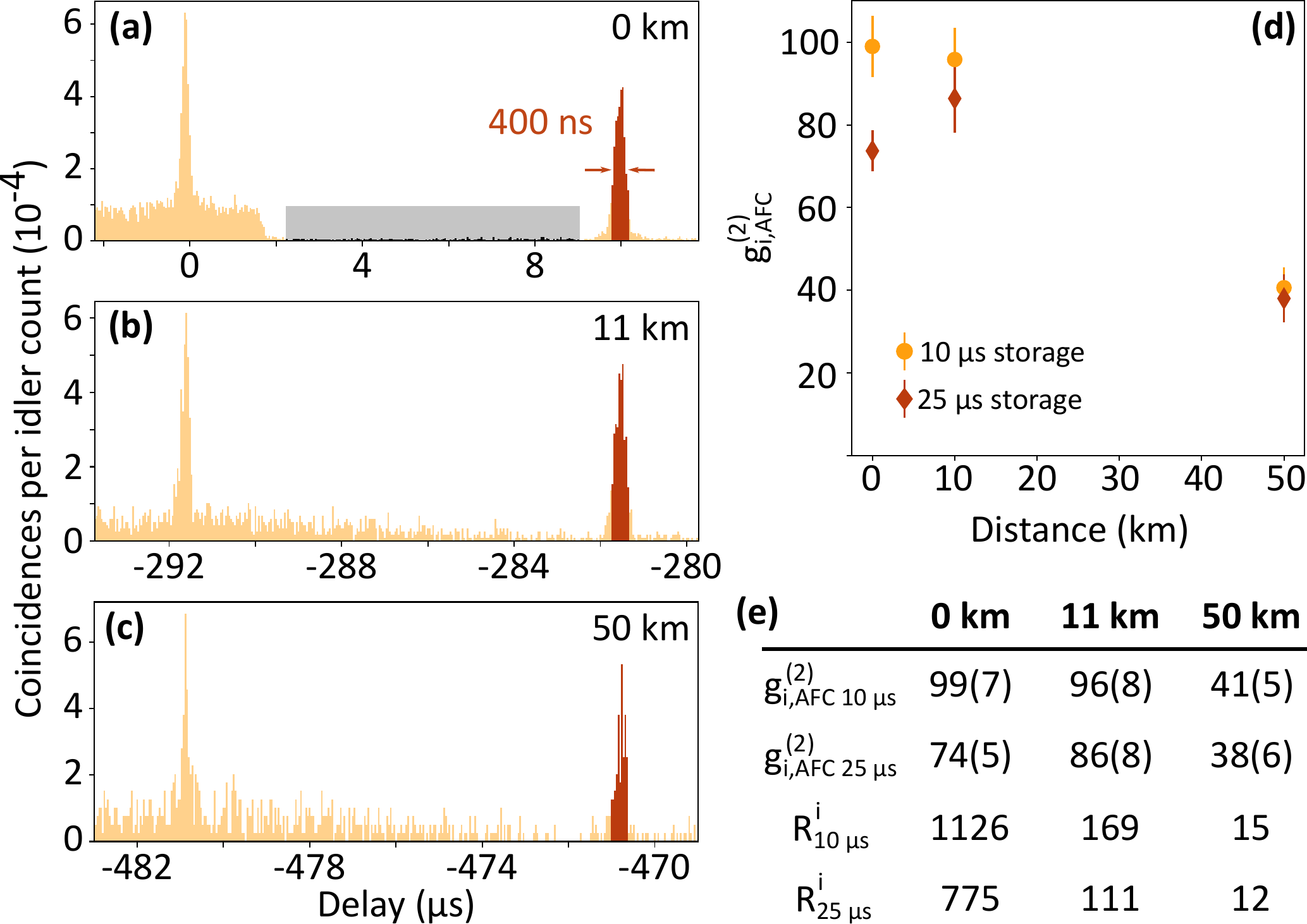}
    \caption{Measurement of cross-correlation. (a-b-c) Coincidence histograms between the retrieved signal photons, stored for 10 $\mu$s, and the telecom idler photons, traveling over an increasing distance. The darker regions are the 400~ns selected for the analysis. In panel (a) it is possible to see the background noise from the SPDC pump disappearing as it is turned off. The grey region is the one selected for the accidental calculation, with counts reported in black. In panels (b)-(c) the triangular pattern of noise is visible. The noise regions are placed in the same position as the echo, but at following periods of the noise pattern (i.e., every 20~$\mu$s after the echo). The integration time for the three histograms were 6, 12 and 42 minutes, respectively. (d) Cross-correlation for all storage times for increasing $d_\tx{idler}$. (e) Table reporting the values of cross-correlation $g^{(2)}_\textup{i,AFC}$ and idler detection rate $R^{i}$ for different storage times and distances. The rates are reported in counts/s.}
    \label{fig:g2}
\end{figure}
\begin{figure}[t]
    \centering
    \includegraphics[width=\columnwidth, left]{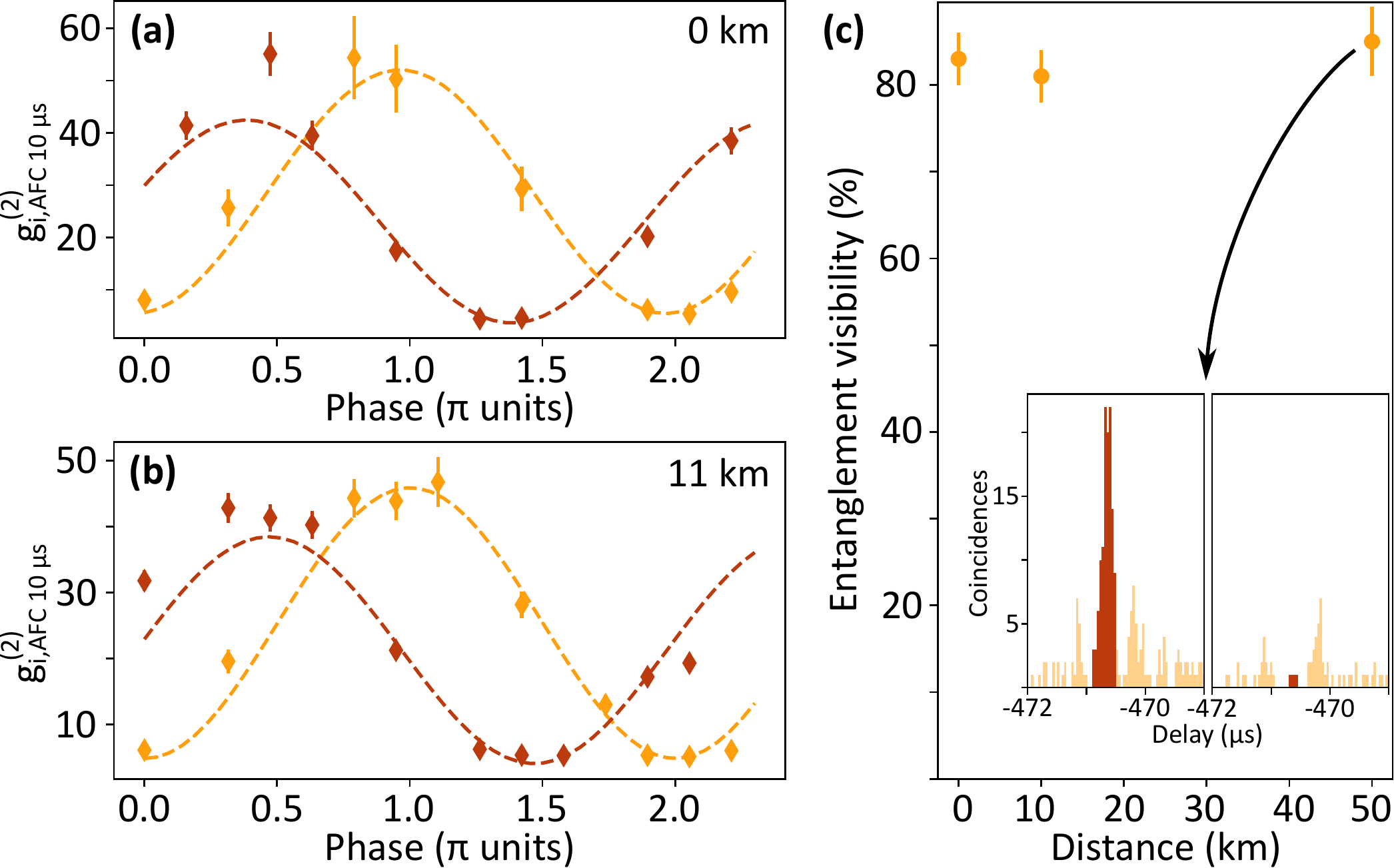}
    \caption{Measurement of light-matter entanglement. (a-b) Interference fringes for 0 and 10 km distance. Each fringe is acquired by changing the phase of the idler interferometer with the fibre stretcher, and between fringes the phase of the signal interferometer is changed by $\pi/2$. (c) Visibility of interference for all $d_\tx{idler}$. The integration times to obtain the three points were 440, 1254 and 1429 minutes, respectively. Inset: histograms of coincidences for $d_\tx{idler} = 50$ km for the maximum and minimum of interference for one of the phase settings of the signal interferometer. The dark regions are those selected for the analysis.}
    \label{fig:vis}
\end{figure}

The results are visible in Fig.~\ref{fig:g2}, where we report the measurement of the cross-correlation $g^{(2)}_\textup{i,AFC}$ for a detection window $\Delta t = 400$~ns. The value of $p_\tx{i}\, p_\tx{s}$ can be estimated by considering the accidental coincidence counts in a region without correlation. The histograms of coincidences for 10 ~$\mu$s storage and for increasing $d_\tx{idler}$ are reported in Fig.~\ref{fig:g2}(a-c) where we highlighted the chosen region for accidentals, and the triangular pattern of accidental counts for higher distances. As visible from Fig.~\ref{fig:g2}(d), which summarises the results at all distances and for both storage times, all values of $g^{(2)}_\textup{i,AFC}$ are well-above the limit of 2 for classical correlations even after the longest fibre link, assuming thermal statistics for the signal and idler fields \cite{Seri2017}. The decrease in the measured value of cross-correlation for the last case can be ascribed to the reduced signal-to-noise ratio of this configuration, due to the higher losses in the idler channel which lower the idler count rate $R^i$. These are reported in Fig.~\ref{fig:g2}(e) and the difference between the two sets is due to the longer preparation time required for an AFC of longer storage time.

\begin{figure*}[t!]
    \centering
    \includegraphics[width=2\columnwidth, center]{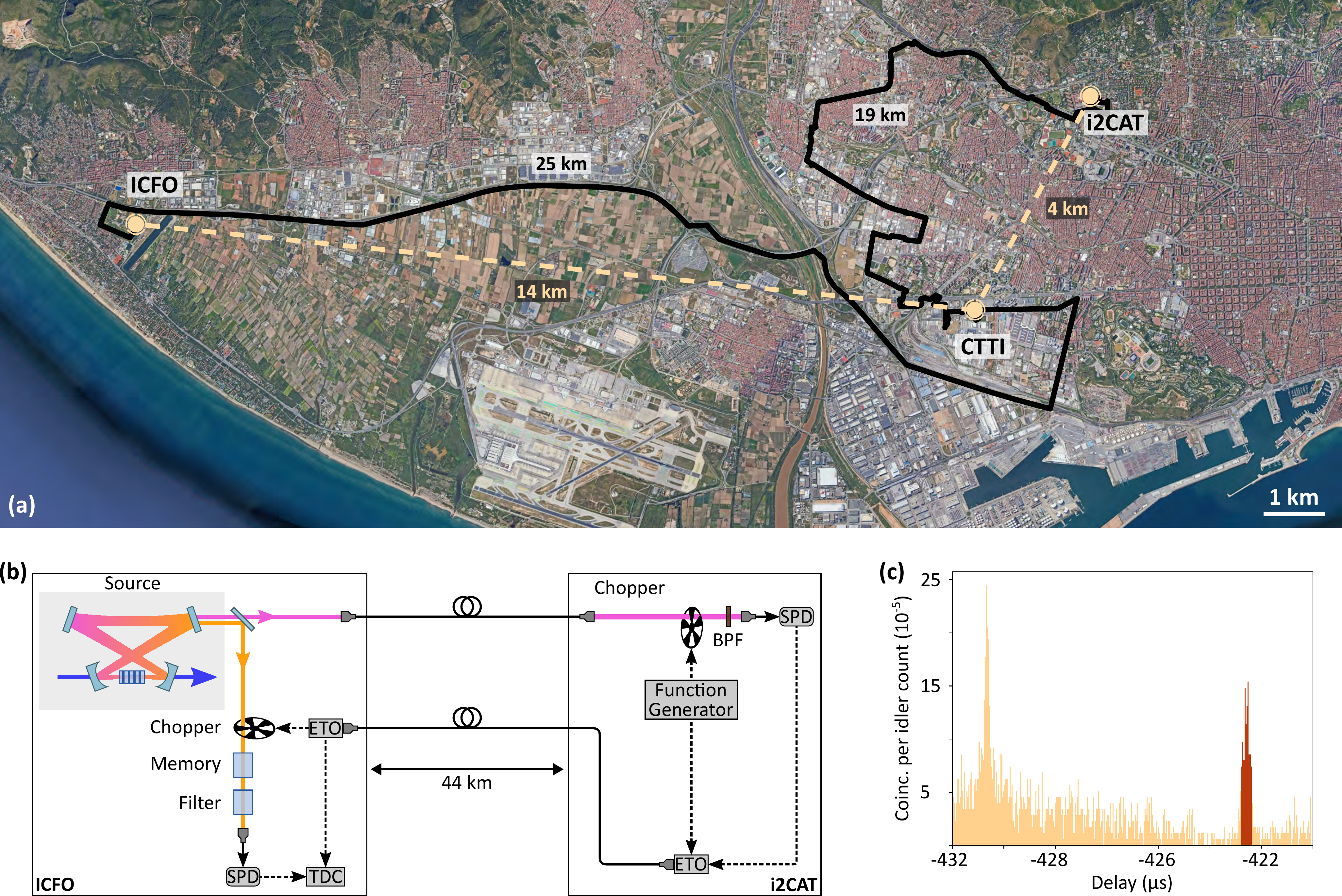}
    \caption{Measurement of non-classical correlations between remote locations. (a) Map of the metropolitan area of Barcelona, with the three locations highlighted: ICFO, where the memory and SPDC source are located; CTTI, where the two optical fibre segments are connected; i2CAT, where the idler photons are detected. Taken from Google Earth (Data SIO, NOAA, U.S. Navy, NGA, GEBCO. Image \copyright\ 2023 TerraMetrics) (b) Schematic of the setup, highlighting the use of the ETO transducers to trasmit the timestamps of the idler detections and the synchronization signal for the mechanical choppers. (c) Cross-correlation histogram built from the TDC at ICFO. Data acquisition was completed in 861~minutes.}
    \label{fig:i2CAT}
\end{figure*}

We then moved to the verification of the entanglement between the signal photon, stored in the multimode quantum memory, and the telecom idler. We employed the Franson scheme \cite{Franson1989}, where we placed an unbalanced Mach-Zehnder interferometer in the path of each photon to act as time-bin analyzer.
This selects two time-bins for the detection of the entangled photon pair, postselecting the joint state $\left ( \ket{ee} + e^{i\varphi} \ket{ll} \right )/\sqrt{2} $, where $\ket{e}$ and $\ket{l}$ are time bins associated with the short and long arm of the interferometer, respectively, and $\varphi$ depends on the relative phase between the two arms. On the idler side we used a fibre-based interferometer \cite{Rakonjac2021} and a solid-state equivalent on the signal side \cite{Clausen2011, Rakonjac2021}. This was realised by preparing in the analysis crystal an AFC with a storage time equal to the delay introduced by the telecom fibre interferometer. The short and long arm are represented by transmission through the AFC and by absorption then reemission, respectively. The phase of this interferometer can be controlled by varying the position of the spectral features of the AFC with respect to the input photon frequency. The phase of the fibre interferometer is instead controlled using a fibre stretcher installed in the long arm. We acquired signal-idler coincidences at various phases of the idler interferometer obtaining a two-photon interference fringe. We then shifted the phase setting of the signal interferometer, acquired a new fringe, and calculated a final weighted average value between the two.
We repeated this process for a $\tau = 10\, \mu$s storage time and for all $d_\tx{idler}$, adopting the same synchronisation between pump and idler recording and the electrical-to-optical signalling explained before. When the metropolitan links were used, the visibility was calculated only by measuring the maximum and minimum values of the two fringes, due to the low count rate. A final value of visibility $V$ higher than 33\% confirms the non-separability of the state of the idler and signal \cite{Peres1996}, while a visibility higher than 70.7\% is enough to violate a CHSH inequality \cite{Clauser1969}. For entanglement-based QKD, a visibility above $78\%$ is required to achieve positive rate \cite{Neumann2021} when finite-size effects are neglected. Our results are reported in Fig.~\ref{fig:vis}, with $V_\tx{0 km}$ = 83(3)\%, $V_\tx{10 km}$ = 81(3)\% and $V_\tx{50 km}$ = 84(4)\%. The values of visibility are all consistent with each other and well above both classical limits for all the distances considered. This is a different behaviour from the cross-correlation, which decreased with increasing distance due to higher losses. The visibility of the interference fringes is affected by signal-to-noise only as $\frac{g^{(2)} -1}{g^{(2)} + 1}$, which remains fairly constant at all distances, and they are in this case mostly affected by the coherence time of the pump laser of the source \cite{Rakonjac2021}. From the measured visibilities we can infer a post-selected two-qubit fidelity of the pair after storage and retrieval of the signal photon in the quantum memory with respect to a maximally entangled state as $F=(3V+1)/4$ \cite{deRiedmatten2005}. For the various distances we obtained $F_\tx{0 km}$ = 87(2)\%, $F_\tx{10 km}$ = 86(2)\% and $F_\tx{50 km}$ = 88(3)\%.

All the measurements mentioned so far have been realised locally, with all the photons being ultimately detected in the same laboratory. We took a step further towards a field-deployed system by bringing the idler detection station to a separate location. We extended the fibre link to an office building of the i2CAT foundation in the city of Barcelona, 17~km away from ICFO and separated by 47~km of fibre. We used one of the two metropolitan fibres to send the idler photons from ICFO to Barcelona, while the other one was used for the transmission of the classical signal of the photon detection. Since the system of superconducting detectors could not be easily moved, we brought a more compact InGaAs detector. This detector had a much lower detection efficiency, about 10\%, but similar dark count rate. The full separation of the system required an additional synchronisation, since one of the optical choppers was now placed at i2CAT, on a portable and blacked-out free-space setup, and it had to be synchronised with the one at ICFO. We therefore placed at the remote location the function generator outputting the master signal that all the choppers used as reference and we transmitted it back to ICFO using another available channel of the ETO transducer. This new fibre link introduced losses of about 13~dB. However the less efficient detector drastically reduced the count rate of the experiment, making the entanglement verification measurement unpractical. We therefore focused on the measurement of the $g^{(2)}_\textup{i,AFC}$ for a storage time of $\tau = 10\, \mu$s. The result is reported in Fig~\ref{fig:i2CAT}(c), where the peak of coincidences is visible among the noise. The lower value of $g^{(2)}_\textup{i,AFC} = 23(2)$ is mostly due to the lower signal-to-noise coming from a lower detection efficiency, but nonetheless clearly confirms the non-classical correlations between the two photons, detected 17~km apart and with a delay of 210~$\mu$s.

\section{Conclusion}
In this work we have presented the detection of non-classical correlations and of light-matter entanglement between a visible photon stored as a delocalised excitation in a solid-state multimode quantum memory and a telecom photon traveling through increasing distances of optical fibre before being detected, both in the laboratory and in the installed fiber network. We measured non-classical cross-correlation values for storage times of 10 and 25~$\mu$s and up to a maximum fibre separation of 50~km between the detected photons, with a reduction due to a lower signal-to-noise. We also verified that light-matter entanglement was preserved after the transmission in the fiber, with two-qubit fidelity $>85 \%$ for all distances. Finally, we physically separated the generation and the detection of the telecom idler photon by placing the detector at a remote location, still measuring a non-classical value of cross-correlation. We modified our system to account for a full separation, introducing synchronisation routines and employing a transducer that allowed precise timings to be transferred over optical fibres, taking the first steps towards the realisation of fully independent quantum nodes. To the best of our knowledge, this is the first demonstration of light-matter entanglement distribution with telecom photons and a multimode quantum memory using deployed fibres.

Achieving this ambitious goal would entail additional requirements, for which our system is well-suited. For most of the measurements presented in this paper, the memory was read-out and the signal photon detected much before the telecom photon reached its destination. On the one hand, this does not affect the quantum correlations as these are calculated based on idler-signal coincidence counts at a particular delay. On the other hand, for applications in quantum networks this post-selection should be avoided. To do so, the storage time of the quantum memory should be greater than the communication time between the two parties, roughly 5~$\mu$s/km. Considering the longest storage time reported in this work, entanglement transmission without post-selection could be achieved across a separation of 5~km and storing 62 temporal modes. On-demand retrieval of the stored photon can be obtained by transferring the excitation to a ground spin state \cite{Afzelius2010a,Seri2017,Rakonjac2021}, also allowing for ultra-long storage times \cite{Holzapfel2020,Ma2021,Ortu2022a} with potential to reach  the level of minutes for a Pr-doped crystal \cite{Heinze2013,Hain2022} due to the long coherence time of the hyperfine levels. On-demand operation is particularly useful when several network links are present, avoiding the necessity of waiting for all of them to be entangled at the exact same time.
The rate of detection of coincidences can also be boosted by improving the efficiency of the memory, and this could be achieved by using an impedance-matched cavity \cite{Afzelius2010b,Jobez2014,Davidson2020,Duranti2023}. Also, in an elementary quantum repeater link \cite{Simon2007,Lago-Rivera2021}, the qubits have to be stored in the quantum memory at least until the heralding photons are detected after propagation in the long fiber and the classical heralding signal come back to the nodes. Therefore we would not be able to turn the source on and off periodically as we did in this experiment, which would significantly reduce the count rate. This limitation can be overcome with a higher level of multiplexing. The Pr-memory used in this work already supported the storage of tens of temporal modes. The GHz-wide inhomogeneous broadening of \PrYSO\ can be exploited for the storage of separate frequency modes \cite{Sinclair2014,Seri2019,Saglamyurek2016,Askarani2021}, and the solid-state crystalline matrix can be designed to exploit spatial multiplexing \cite{Gundogan2012,Yang2018}, making rare earth-doped crystals a promising system for high levels of multiplexing \cite{Ortu2022b}, and an equally promising candidate for the realisation of the basic building block of a future quantum network.

\section{Acknowledgments}
We thank Susana Plascencia for help in the early stages of the experiment. We thank Xavier Jordan Parra and Francisco Miguel Tarzan Lorente of i2CAT for providing the space and the fibre connection to the Nexus 2 building. We also thank Joaquim Garcia Castro of CTTI and Xarxa Obierta de Catalunya for providing access to the metropolitan fibre link.

This project received funding from the European Union Horizon 2020 research and innovation program within the Flagship on Quantum Technologies through grant 820445 (QIA) and under the Marie Sk\l odowska-Curie grant agreement No. 713729 (ICFOStepstone 2) and No. 754510 (proBIST), from the European Union Regional Development Fund within the framework of the ERDF Operational Program of Catalonia 2014-2020 (Quantum CAT), from the Government of Spain (PID2019-106850RB-I00; Severo Ochoa CEX2019-000910-S; BES-2017-082464), from the MCIN with funding from European Union NextGenerationEU PRTR (PRTR-C17.I1; MCIN/AEI/10.13039/501100011033: PLEC2021-007669 QNetworks; IJC2020-044956-I Juan de la Cierva Fellowship to S.G.), from Fundaci\'o Cellex, Fundaci\'o Mir-Puig, and from Generalitat de Catalunya (CERCA, AGAUR).

%


\end{document}